# Blockchain Technology: A tool to solve the challenges of education sector in developing countries

Md Aminul Islam

Faculty of Engineering, Technology and Computing [ECM]

Oxford Brookes University, OX3 0BP, UK

Email: 19126681@brookes.ac.uk, talukder.rana.13@gmail.com

## Abstract

The education system is getting diversified, challenged, and blended for the overwhelming advancement of disruptive technology. The core purpose of this chapter is to visualize the probable solutions of the modern education system using blockchain technology. The entire chapter has been discussed on the basis of present solution and projection of future inventions to smoothen the education system. The fourth industrial revolution (4IR) is changing our experiences in terms of education and other lifestyle. Delivering lectures, interacting between learners and educations, evaluating learning outcomes, and verifying educational credentials might be smoother, easier, faster, cheaper, and jollier than before. Blockchain technology can contribute to the education provider to tackle all those existing problems to create a comfortable learning environment to all irrespective to their economic backgrounds and geographic location. How this technology can contribute to improve Reviewing recent inventions in this technology, the chapter explains some of the strategies to go beyond the ongoing projects around the world. A set of models are arranged to enable the readers mind for future inventions in the realm of educationists.

# 1. Introduction

## 1.1 Background:

In 2008, a person going by the name Satoshi Nakamoto proposed using blockchain technology for virtual currency. Nakamoto developed bitcoin, a digital money based on the blockchain. This network stores and disseminates data using distributed ledger technology and operates on a peer-to-peer model (Nakamoto, 2009). Numerous alternative cryptocurrencies appeared on the market after Satoshi Nakamoto used blockchain technology to create bitcoin. Essentially, blockchain is a shared ledger that records and verifies financial dealings between parties. To put it simply, blockchain is a distributed database that records digital transactions. The blockchain is a distributed ledger in which each user's transaction history is recorded in chronological order and could be altered by a consensus of all users (Sarmah, 2018). Private blockchains are exclusively utilised internally at a corporation or organisation, while public blockchains are available to the general public. There are several different blockchains, but the one at the heart of Bitcoin's cryptocurrency is the most well-known. Various applications exist for DLT's tamper-proof transactions and its ability to maintain a clear and unambiguous register of information exchange (Ark, 2018).

To address the need for decentralisation and confidentiality of cycles common to ordinary people and organisations, the blockchain is a large and global encoded data set that puts an end to the distribution or monopolisation of data by the parts that interacted with that data. An enormous, decentralised, encoded, and open book of records in which experts influence the veracity of the data and guarantee its moral correctness were discussed. Applications of this type of spectacular and general arrangement include the exchange of decryption forms of money, decentralisedprogramme distribution, and record keeping (Gomez, 2021).

At first, blockchain technology was mainly used for digital currency transactions, but now there is a growing body of research on the potential benefits of blockchain in other industries, including medicine, education, urban planning, finance, insurance, and many more. The blockchain provides an impregnable and unhackable network for computer-based applications, therefore the trend has turned towards the secure and verified system (Sharples, & Domingue, 2016).

Now, in terms of the education industry, the rapid development of distributed information and blockchain technology has prompted us to reassess and reexamine several fundamental components of existing education, literacy, and training frameworks. With the introduction of this new system of improvements, previously held beliefs about things like trust, value, security, and character are being called into question. Because of the fast development of distributed computation and blockchain technologies, many of our long-standing educational institutions are being rethought and redesigned. When blockchain technology is introduced to the classroom, it raises new questions about the nature of ideas like trust, privacy, and identification, as well as a whole new set of technologies (Sharples, & Domingue, 2016). The primary focus of blockchain research in the education sector is on the ways in which the technology might facilitate the safe, reliable, and auditable dissemination of knowledge. The ledger then serves as a central repository for all of the related educational institutions to access and use in the course of their respective teaching, learning, and accrediting processes (Bhaskar et al., 2020). Researchers have discovered that Blockchain technology creates a setting where students may act as their own registrars and where third parties are not required to record or modify their grades. Education providers may also use blockchain technology, a decentralized data exchange, to issue, validate, and share certificates, which will assist to reduce the prevalence of certificate fraud (Agarwal et al., 2021).

The education industry has been highlighted as a revolutionary arena for technology, a rising area where the credential is vital. Multiple studies have indicated that an appreciable number of candidates forged their academic credentials. It is not coincidental that we might benefit from avoiding certain potential arguments by double-checking this kind of data now (Sharma, 2018). The education industry relies heavily on student data, and blockchain technology has the potential to make resources like student enrollment, tuition paid, course completion, and even a student's diploma an integral part of their digital identity. Information security is improved by the immutability of these records. Rather than belonging to the institution where they were created, students should retain ownership of their academic records. Once a record is recorded on a blockchain, no user may change it. In the unlikely event that a record contains an error, a new record should be created to correct it, and both the incorrect and correct records will remain visible (Hance et al., 2021).

This whole chapter has been written with the aforementioned problems in mind, discussing them in light of existing solutions and anticipating future innovations that will further streamline the educational process. This chapter examines recent developments in this field of blockchain technology and outlines some of the methods that may be used to go surpassing the already active initiatives in the globe.

## 1.2 Aim of Research

Studying blockchain's potential as a tool to address issues plaguing the global education system is fundamental to this study. In particularly, the specific objectives are to:

1. Classify the development of research areas concerning the blockchain
2. Identify the various blockchain applications and categorise them.
3. Evaluate the practicality of blockchain technology.
4. Separate the current problems with contemporary education in underdeveloped countries from the challenges and opportunities of implementing blockchain technology in education, and
5. Connect the dots between the discovered advantages and threats of blockchain implementation in education and the identified research topics.

## 1.3 Research Question:

RQ1: How can the educational system be integrated with Blockchain technology?

RQ2: What are the implications of Blockchain for the education systems of poor nations?

RQ3: What are the challenges to adopt Blockchain technology for education system?

## 1.4 Methodology:

In order to address the research objectives posed by the study, a thorough content analysis of the available literature has been conducted as a secondary data source. The purpose of content analysis is to identify recurring ideas, topics, and terminology within a body of qualitative data (typically text). Researchers may use content analysis to determine the frequency with which certain words, ideas, or concepts appear in a text, as well as their significance and interrelationships. Investigators may check articles for signs of prejudice or partiality by analysing the words and phrases used. This allows them to draw conclusions about the book's meaning, the author, the intended readers, and the historical and cultural context in which the piece was written (Harwood and Garry, 2003). Content analysis may be broken down into two broad categories: conceptual analysis and relationship analysis. Using this method, we may ascertain whether or not a text contains any ideas and how often they appear. By digging further into the interconnections between ideas in a text, relational analysis expands on the conceptual analysis. Results, conclusions, interpretations, and meanings may vary depending on the method used to analyse the data. Conceptual analysis was performed with content analysis in this study (Lacy et al., 2015).

Books, journal articles, essays, talks, newspaper headlines, lectures, media, historical records, and online forums are all mined for information for this study. The study used a process called content analysis, in which the text was dissected from several perspectives to reach its goal. The primary method used for this was a literature of previous research. A literature review is an approach of collecting, analysing, and interpreting the research that has been done on a certain subject, phenomena, or research issue (Snyder, 2019). Primary studies are the individual study projects that make up a systematic review, which is itself a subset of secondary research. Following the guidelines established by Petersen et al. (2008), we performed a search for applicable publications and compiled them into a literature review. Because this content analysis review set out to discover pre-existing research on Blockchain technology, the review procedure was used as the study approach.

# 2. Blockchain: The associated issues with the technology features

## 2.1 Blockchain technology

Bitcoin, Ethereum, and other cryptocurrencies all use blockchain, a new kind of database technology. Blockchain prevents hacking and fraud by dispersing several, identical databases throughout a network. Bitcoin and other cryptocurrencies may be the most visible use of blockchain right now, but the technology has the potential to benefit a broad variety of industries. Blockchain may be thought of as a decentralised digital ledger that can be used to record and verify transactions of any sort. Non-fungible tokens and cryptocurrency transactions may both be recorded on a blockchain (Yli-Huumo et al., 2019). While this data may be stored in almost any database, blockchain is special because it is distributed. In contrast to centralised databases like Excel spreadsheets or bank ledgers, which are stored on a single server, blockchain databases are distributed throughout a network of computers. Nodes are the individual computers that make up a network (Sheth and Dattani, 2019).

## 2.2 How Blockchain works

In many ways, blockchain technology is more important than the internet itself. There is no need for credibility or a governing body to facilitate the transfer of wealth (Yaga et al., 2019). Explaining the inner workings of blockchain, or distributed ledger technology, using Bitcoin as an example in here:

1. Bitcoin transactions are recorded and sent to a series of supercomputers, or nodes, which verify and process the trades.
2. Using complex mathematical procedures, hundreds of nodes in the network compete to verify the transaction. Bitcoin mining refers to this process. The miner who finishes a block first receives Bitcoin as payment for their efforts. These bonuses are paid for using freshly generated Bitcoin and shared network fees. Transaction volume will determine the range of transaction fees.
3. The transaction is added to the distributed ledger as a new block after the cryptographic confirmation of the purchase has been received. It is then up to the vast bulk of the network to approve the transaction.
4. A cryptographic fingerprint called a hash is used to link the block to all preceding block of Bitcoin transactions, and the transaction is finalised (Niranjanamurthy et al., 2019 ).

.

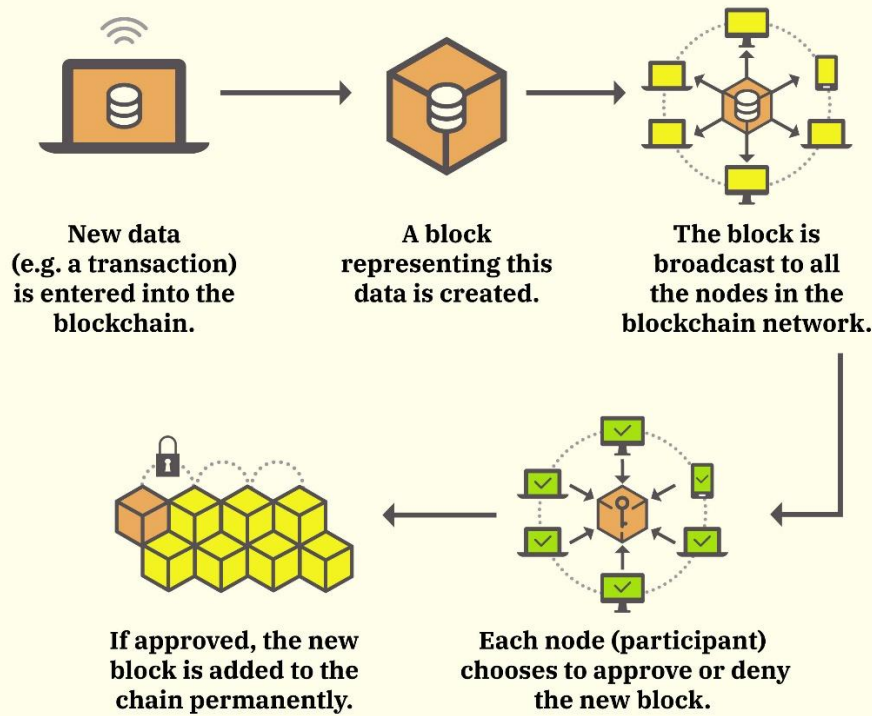

Figure: Steps of how blockchain works (Fox, 2017)

## 2.3 Different Types of Blockchain Technologies

There are four primary categories of blockchain systems to consider. The following are cases (Wegrzyn and Wang, 2021):

**Public Blockchain**

One implementation of blockchain technology is a public blockchain. If you are looking for a distributed ledger system that anybody can participate in and conduct transactions on, look no further than a public blockchain. This approach imposes no limits by having each peer maintain their own edition of the ledger. This further implies that anybody with an internet connection may get entry to the public blockchain. The Bitcoin block chain technology was among the first distributed ledgers to be made available to the general public. It made it possible for any web user to conduct decentralised financial dealings.

**Private Blockchain**

One use of blockchain technology is a private blockchain, which is only accessible to a select group of users. The term "private blockchain" is often used to refer to a blockchain that operates in a closed network. It is likewise a blockchain where access is restricted to those who have been granted permission to use it. A private blockchain is ideal for a corporation or organisation that wants to utilise the technology internally but keeps its operations secret. This way, you may restrict access to the blockchain network to a restricted group of users while still reaping the benefits of blockchain technology. A variety of network settings, such as access, authorisation, and so on, may be adjusted by the organisation.

**Consortium Blockchain**

One implementation of blockchain technology is a "consortium blockchain." Consortium blockchains, sometimes called Federated blockchains, are an innovative solution for businesses who need both public plus private blockchain capabilities to meet their demands. A consortium blockchain allows for both public and private information to be stored and shared. The consortium blockchain's consensus processes are managed by the predetermined nodes. Moreover, it maintains its decentralised character while being inaccessible to the general public. How? First off, a consortium blockchain is overseen by many entities. In other words, the results are not being governed by a central authority.

**Hybrid Blockchain**

One kind of blockchain technology is called a "hybrid blockchain." Last but not least, we will talk about a blockchain architecture known as a hybrid blockchain. And although a consortium blockchain may seem similar, that is not what a hybrid blockchain is. On the other hand, there may be some overlap. A hybrid blockchain is a blockchain that combines private and public elements. The technology has applications in businesses that are not committed to either private or public blockchains but nevertheless want to take use of the advantages of both.

## 2.4 Use of blockchain technologies in everyday life.

Everyday applications of blockchain technology range from commercial transactions to the management of electoral processes. The following are some primary usages based on the regular life:

**Cryptocurrency**

Blockchain is most widely used today as the underlying technology for digital currencies like Bitcoin and Ethereum. The activities of Australians who purchase, trade, or spend cryptocurrencies are all recorded on a distributed ledger called a blockchain. Blockchain's potential for wider adoption increases as bitcoin use rises. Cryptocurrencies are not widely utilised for making purchases because of their high volatility. However, this is beginning to change as digital asset services are becoming more widely accessible to merchants and retail consumers via the likes of PayPal, Square, and other payment service providers.

**Banking**

Blockchain is also being used to execute transactions with fiat currencies such as the pound, the euro, and the eurozone's equivalent of the franc. If the payments can be confirmed and handled outside of typical business hours, this might be a speedier option than transferring money via a bank or financial institution.

**Asset Transfers**

As with the transfer of funds, blockchain technology may be used to keep track of and transfer ownership of other assets. This is especially common with NFTs, a depiction of ownership used for digital artwork and video, which has recently gained in popularity. But blockchain technology might also be employed to record the legal transfer of physical goods like titles to homes and cars. As a first step, the blockchain would be used to confirm that both parties have the funds and title to the property. If everything went well, the blockchain could be finalised and recorded on the distributed ledger. If they used this method, the deeds to the property could be transferred without any intermediaries (such as a notary public) being required since the transaction would be recorded on the blockchain immediately after it took place.

**Smart Contracts**

Another blockchain breakthrough is self-executing contracts generally termed "smart contracts." When certain criteria are satisfied, these electronic contracts are immediately put into action. For instance, a purchase for an item may be issued quickly if the sellers and buyers have completed all set requirements for a trade.

**Supply Chain Monitoring**

Since items go from one region of the globe to another, supply networks generate enormous volumes of data. Identifying issues, such as the manufacturer of low-quality items, may be challenging if data is stored in the conventional manner. IBM's Food Trust employs blockchain system to detect food from it's own harvest to its consumption, and storing this knowledge on blockchain would allow it easy to go back and check the supply chain.

**Voting**

A group of experts is also investigating how blockchain technology may be used to reduce voter fraud. In principle, blockchain voting would do away with the need to physically collect and validate paper ballots and would also enable voters to cast votes that are impossible to tamper with.

**Education**

During the epidemic, educational institutions jumped on the digitising bandwagon. As a result of the revolutionary nature of blockchain technology, this industry may see a dramatic shift. To begin, blockchain has the potential to drastically improve the ways in which academic collaboration and record keeping are handled. Since blockchain is a distributed ledger, it has the potential to greatly improve the sector by increasing openness and responsibility in technology. Let us take a look at how blockchain technology might impact classroom instruction (Lindman et al., 2017; Purohit, 2022).

# 3. Challenges of modern education

## 3.1 Challenges of traditional education system

The following challenges are accumulated by the information from on Stovall (2005) and Chiţiba (2012):

**Less learning Outcome:**

There are not enough learning outcome sets to blame for the collapse of the old school system. Most students seldom stop to consider why they are studying something or what they hope to gain from doing so. That begs the question, "Why have we started this course now?" The fundamentals of beginning a chapter with questions regarding what I will gain from reading this chapter are never emphasised. The only time students read the chapter is just before an exam. Their main concern is doing well on the test. They consistently show a lack of study in what they should be learning.

**Learning level is not the same:**

Those attending a class come from a wide variety of origins. It is likely that their parents' schooling experiences were unique. Parental academic attainment is not often high. While a handful of the students will be first-generation study-goers, the majority are third-generation students. Thus, the third-generation student's parental participation in their child's schooling is superior to that of the first-generation learner.

**Lack of skill-based education:**

When compared to more modern forms of education, the conventional one places more emphasis on imparting generalised information rather than practical training. Innovation, originality, and "out of the box" thinking are often discussed in the business world. However, none of them are supported by the mainstream educational system. We live in a democracy where leadership can be changed, but our system of education is not built to foster growth over the course of years.

**Non-availability of quality teachers:**

In conventional classrooms, teachers are of subpar quality. People often see teaching as a career of last resort. Teachers, even those who are formally trained in the field, often fail to pass on useful knowledge to their students. Commercial interests have penetrated the academic realm. Earning more money or landing a cushy job has replaced helping students learn as the main purpose of teaching.

## 3.2 Challenges of modern and digital education system

1. Sadly, few educational institutions have the technological wherewithal to provide digital textbooks and reading materials on a wide scale. Digital books demand fast internet, which is not widely available in the poor world. However, developed countries like the US have no trouble with the huge data downloads that this entails.
2. It has never been the technology itself that has stood in the way of progress in this area, but rather the people who have been directly impacted. Everyone in the education system is mired in the past, from classroom instructors to school principals to librarians to parents. The difficulty comes from pushing students to accept digital learning as the norm.
3. The content of digitising educational materials extends much beyond the simple digitization of textbooks and other materials already in existence. Dynamic and engaging curated content is essential for the success of digital learning in the classroom. Curating this kind of content is time-consuming and labor-intensive, which contributes to higher overall implementation costs.
4. The cost of implementing a virtual curriculum is not a fixed amount. As technology evolve, so too must the curriculum reflect these changes. Changes in technology may have profound effects on previously published content (Shutikova and Beshenkov, 2020; Ronzhina et al., 2021).

## 3.3 Challenges of education posed by rapid changes

A lot has changed in the way individuals work during the last 30 years. In modern, industrialised cultures, the practise of workers remaining at the same company for a quarter century is obsolete. As a result, today's schools and universities must equip their pupils to hold ten or more occupations before they are 50. Companies want to hire people who are not just knowledgeable in their fields, but also adaptable, mature, and fast learners (Oyedotun, 2020).

The whole structure of the economy is shifting. Globally, the information economy and the consumer economy are booming. The non-oil economy in Abu Dhabi is expected to increase by more than eight percent annually as the city continues its long-term transformation to a stable, high-value knowledge economy. The knowledge-based sectors of the future will demand people with strong communication, teamwork, and creative abilities. Each person must adopt a new way of thinking in order for globalisation to work. Fifty years ago, the globe seemed much larger than it does now. In order to appreciate the intricacies of other people's cultures, our pupils will need to be "globally competent." Information has also undergone profound changes in the modern era. Every two years, the total quantity of data in the globe doubles. Therefore, we will have to adjust our approach to this management. In ways we cannot foresee at the moment, students of the tomorrow will need to understand how to filter, collect, and synthesise data (Nurhas et al., 2021).

# 4. Blockchain in education

## 4.1 Use of blockchain in education

During the epidemic, educational institutions jumped on the digitising bandwagon. As a result of the revolutionary nature of blockchain technology, this industry may see a dramatic shift. To begin, blockchain has the potential to drastically improve the ways in which academic collaboration and record keeping are handled. Since blockchain is a distributed ledger, it has the potential to greatly improve the sector by increasing openness and responsibility in technology. As the globe becomes more technologically sophisticated, the educational system stands to be shaken up. The education technology sector has benefited us for twenty years. It is safe to say that this trend has hastened the process of bringing schools up to date. Now is the moment for blockchain technology to greatly quicken the process. The distributed database of the blockchain, AI, and machine learning are gradually displacing textbooks (Maryville, 2021). Below, we highlight the applications and consequences of blockchain technology in the academic sector (Bhaskar et al., 2020; Rahardja et al., 2020).

**Intelligent Agreements for Courses and Assignments**

Agreements are often implemented on blockchains. This may help instructors create blockchain-based courses and lessons. After the requirements have been met, the class will be taught automatically and at the student's own speed. Students and instructors might sign a contract detailing assignment restrictions, due date, and grading deadline.

**Certifications, Report Cards, And Documentation**

The immutable ledger technology of blockchain produces a chronological record of recent occurrences. This might be useful for presenting student transcripts, producing a thorough report card, monitoring attendance, and informing students and stakeholders about their progress. Using blockchain, students may submit homework without concern of losing them. Additionally, students may obtain their degrees and certificates online, as opposed to on fragile paper. Digital degrees & certificates are favoured since they are hassle-free, well-organized, and uncomplicated.

**Streamlining the Payments of Fee**

The procedure of paying student tuition is time-consuming and difficult. Students, parents, banks, organizations or government organisations for grants, lenders, and other university departments are involved. This process, however, can be expedited using blockchain technology, resulting in decreased administrative costs and maybe even cheaper tuition fees.

**Universal Admittanceand Lower Expense**

To encourage and facilitate lifelong learning, blockchain technology may facilitate the distribution of freely available, open educational content like books, lectures, and films in the public domain. Blockchain technology allows for the safe and low-cost public sharing of such assets. Additionally, blockchain enables teachers to assess their students' work on the blockchain, making it possible for students in far-flung regions to participate in digital versions of courses and exams.

## 4.2 How blockchain can solve the challenges of education system

According to Wallace (2019) and Steiu (2020), below are some of the sectors of the education business where Blockchain might have an influence and how it can tackle the issues of the education system:

- The purpose of higher education is dual in the field of research: first, to preserve information for future generations of students, and second, to increase current knowledge via research. Professors spend a great deal of time performing research work and publishing their results, expanding the boundaries of their disciplines and uncovering new areas of study that will lead us into the future. Moreover, the reach and effect of such articles may have an influence on academics' potential to seek large funding to cover future research. There is a strong incentive for authors to monitor the distribution of their works and take measures to discourage blatant theft.
- Researchers would be free to publish anything they choose, so long as they publish in a fashion that allows them to measure the reuse of their work (such as how frequently it is referenced or used as teaching materials), which is crucial since it may lead to recognition and further funding.
- While still in its infancy, blockchain in E - learning has been heralded as "the ideal technology" for retooling an antiquated system.
- With the spotlight shining brighter than ever before, Blockchain's distributed ledger keeps technology of transactions in near real-time and cannot be altered once they have been recorded. This is a great way to maintain tabs on kids' academic standing, show a full report card, and verify transcripts. Students who use blockchain to turn in their work cannot complain that their submission was misplaced.
- Smart contracts may also be used to emphasise responsibility. Teachers, administrators, and students at educational institutions will soon have access to smart contracts. To clarify the parameters of an assignment, the due date, and the grading schedule, for instance, students and professors might sign a digital agreement. It is possible that student loan debt may be settled via smart contracts.
— When applied to the problem of reducing the price of higher education, it has the potential to have a dramatic effect. Open educational materials are those that are in the digital realm and free to use and redistribute, and they might be made accessible to everyone via the use of blockchain technology, in addition to their potential utility in promoting lifelong learning. Blockchain technology makes it possible to safely and cheaply distribute these kinds of resources on a public network.
— Blockchain has the ability to change the education landscape because it can address these issues by providing new, less expensive means of teaching and by breaking the current link between schools and students (Bhaskar et al., 2020; Agarwal et al., 2021).

## 4.3 What are the current and possible limitations in adapting blockchain in education sector

Blockchain has a lot of promising applications, but it has not yet seen widespread adoption. More than half of the students polled by Gartner said they had no plans to use blockchain technology. The difficulties of putting into practise the technology might be to blame for much of the resistance (Alammary et al., 2019).

**Security**

Even while blockchain's security is a major selling point, it is not bulletproof. Institutions need to be careful about what material they save and how they select to store it due to the sensitive nature of the information being kept on the blockchain (students' academic records and credentials). There may also be difficulties associated with meeting the requirements of federal and state privacy legislation. More stringent privacy safeguards may be required at universities, such as the use of private or private blockchain or the encryption of blockchain-stored data.

**Scalability**

Large amounts of student and alum data held by educational institutions might provide a scaling challenge for blockchain applications. More blocks are needed to accommodate more data, which slows down blockchain transactions since each one must be verified by the network's nodes. This may be a serious barrier to widespread implementation. A benefit of permissioned blockchains is that they can process more transactions per second than permissionless ones.

**Adoption rate**

Blockchain, like other technology before it, is only useful when enough universities and employers depend on it; students only gain from proprietorship of their diplomas if the schools or firms to which they are applying recognise their validity. Many job boards, like Upwork and

ZipRecruiter, are actively encouraging blockchain-based credentials, and hundreds of colleges are now issuing and accepting them.

**Cost**

Adopting and deploying any new technology may be fairly expensive, despite the fact that it might lead to benefits in other areas. Investing in more computer resources or upgrading a current infrastructure may be expensive. Institutions may also need to spend time and money training school administrators to utilise the technology (Steiu, 2020; Lutfiani et al., 2021; Anand, 2022). This is because many organisations may lack the expertise and skills essential to handle student data on a public blockchain.

### 4.4 Context of Models Suggested for Developing Countries

Regarding the view from underdeveloped nations According to UNESCO's definition, open educational resources are those that are either freely available in the public domain or licenced in a way that allows for their reuse, modification, and redistribution by anyone. Open textbooks, courses, and curricula are only the beginning of what may be found in the realm of free and open educational materials. It is not limited to textbooks, though; software, podcasts, and movies may all serve the same purpose. They cut the price of educational materials for pupils by a significant content. Teachers and students have been given more control over their learning with the use of these tools since they have faster and easier access to more high-quality content. Taking this definition into account, Blockchain may be thought of as a decentralised digital ledger as well as database operating on a network. Because the technology is decentralised, open educational materials (the "blocks" in the "chain") may be safely and efficiently shared across users on the internet. Blockchain technology can facilitate the worldwide distribution of free and open educational content. Blockchain was first developed for use in Bitcoin as a distributed ledger to record encrypted and verified financial transactions between users over a network ( McGreal, 2021).

For example, look at the Blockchain for Education Community of Practice from the World Bank (COP). It is a hub for anyone interested in using blockchain for positive change in educational institutions. Members offer insights gained through their research and experimentation into the reality of making the most of connected technology. Educators and policymakers in charge of public school systems who are interested in enhancing their classrooms with technological resources are a key target demographic for this COP. This strategy may be adopted in underdeveloped countries with the help of the World Bank's EdTech Team, wherein participating organisations and Ministries of Education work together to conduct pilots and participate to accessible global public goods (World Bank, 2021).

## 5. Concluding Remarks and Way Forward

The application of blockchain technology in education is still in its infancy. In order to identify specific gaps in research that must be considered in future studies, it is required to conduct a study of current blockchain work in the domain of education. This study aimed to perform a thorough review of blockchain implications in the education industry. For this purpose, this paper conducted a literature review based on content analysis. Our literature review enabled us to assess the benefits and drawbacks of using blockchain technology in educational arena. Consequently, our research may assist future attempts to overcome the limitations of current solutions in the study of blockchain integration in education. Even though this innovation is still in its infancy and must be validated, it has the capacity to inspire a tremendous lot of future innovation. It will be feasible to discuss the educational uses of blockchain technology in detail in future work. Moreover, the monetisation of all student affairs on a blockchain system will motivate students to participate in a range of activities, including sports, attendance, assignments, etc.

# 6. References

Agarwal, P., Idrees, S.M. and Obaid, A.J., (2021). Blockchain and IoT Technology in Transformation of Education Sector. *International Journal of Online & Biomedical Engineering*, *17*(12).

Alammary, A., Alhazmi, S., Almasri, M. and Gillani, S., (2019). Blockchain-based applications in education: A systematic review. *Applied Sciences*, *9*(12), p.2400.

Anand, A. (2022). *Blockchain in Education Sector: Advantages and Disadvantages | Analytics Steps*. [online] www.analyticssteps.com. Available at: https://www.analyticssteps.com/blogs/blockchain-education-sector-advantages-and-disadvantages[Accessed 14 Nov. 2022].

Ark, T. V. (2018). *20 ways blockchain will Transform (Okay, May IMPROVE) EDUCATION*. Forbes. https:// www.forbes.com/sites/tomvanderark/2018/08/20/26-ways-blockchain-will-transform-ok-may-improve-education/?sh=cb87fff4ac91[Accessed 12 Nov. 2022].

Bhaskar, P., Tiwari, C.K. and Joshi, A., (2020). Blockchain in education management: present and future applications. *Interactive Technology and Smart Education*.

Chițiba, C.A., (2012). Lifelong learning challenges and opportunities for traditional universities. *Procedia-social and behavioral sciences*, *46*, pp.1943-1947.

Fox, G. (2017). *How Blockchain Works Infographic*. [online] GARY FOX. Available at: https://www.garyfox.co/how-blockchain-works-infographic/ [Accessed 15 Nov. 2022].

Gomez, D. (2021). *Blockchain in education, a large and global encrypted database*. eLearn Center Blog. https:// elearncenter.blogs.uoc.edu/blockchain-in-education/[Accessed 09 Nov. 2022].

Hance, M. (2021). *What is blockchain and how can it be used in education?* MDR. https://mdreducation.com/2018/08/20/blockchain-education/[Accessed 10 Nov. 2022].

Harwood, T.G. and Garry, T., (2003). An overview of content analysis. *The marketing review*, *3*(4), pp.479-498.

Lacy, S., Watson, B.R., Riffe, D. and Lovejoy, J., (2015). Issues and best practices in content analysis. *Journalism & mass communication quarterly*, *92*(4), pp.791-811.

Lindman, J., Tuunainen, V.K. and Rossi, M., (2017). Opportunities and risks of Blockchain Technologies–a research agenda.

Lutfiani, N., Aini, Q., Rahardja, U., Wijayanti, L., Nabila, E.A. and Ali, M.I., (2021). Transformation of blockchain and opportunities for education 4.0. *International Journal of Education and Learning*, *3*(3), pp.222-231.

Maryville, O. (2021). *How Blockchain Is Used in Education*. [online] Available at: https://online.maryville.edu/blog/blockchain-in-education/#role [Accessed 11 Nov. 2022].

McGreal, R. (2021). *How blockchain could help the world meet the UN's global goals in higher education*. [online] The Conversation. Available at: https://theconversation.com/how-blockchain-could-help-the-world-meet-the-uns-global-goals-in-higher-education-152885#:~:text=a%20public%20network.- [Accessed 15 Nov. 2022].

Nakamoto, S., 2008. Bitcoin: A peer-to-peer electronic cash system. *Decentralized Business Review*, p.21260.

Niranjanamurthy, M., Nithya, B.N. and Jagannatha, S.J.C.C., (2019). Analysis of Blockchain technology: pros, cons and SWOT. *Cluster Computing*, *22*(6), pp.14743-14757.

Nurhas, I., Aditya, B.R., Jacob, D.W. and Pawlowski, J.M., (2021). Understanding the challenges of rapid digital transformation: the case of COVID-19 pandemic in higher education. *Behaviour & Information Technology*, pp.1-17.

Oyedotun, T.D., (2020). Sudden change of pedagogy in education driven by COVID-19: Perspectives and evaluation from a developing country. *Research in Globalization*, *2*, p.100029.

Petersen, K., Feldt, R., Mujtaba, S., Mattsson, M. (2008). Systematic mapping studies in software engineering. In 12th International Conference on Evaluation and Assessment in Software Engineering (EASE) 12 (pp. 1–10).

Purohit, A. (2022). *5 Ways Blockchain Impacts The Education Industry In 2022 And Beyond*. [online] eLearning Industry. Available at: https://elearningindustry.com/ways-blockchain-impacts-education-industry-in-2022-and-beyond [Accessed 14 Nov. 2022].

Rahardja, U., Aini, Q., Ngadi, M.A., Hardini, M. and Oganda, F.P., (2020), October. The Blockchain Manifesto. In *2020 2nd International Conference on Cybernetics and Intelligent System (ICORIS)* (pp. 1-5). IEEE.

Ronzhina, N., Kondyurina, I., Voronina, A., Igishev, K. and Loginova, N., (2021). Digitalization of modern education: problems and solutions. *International Journal of Emerging Technologies in Learning (iJET)*, *16*(4), pp.122-135.

Sarmah, S.S., (2018). Understanding blockchain technology. *Computer Science and Engineering*, *8*(2), pp.23-29.

Sharma, A. (2018). *Blockchain could revolutionize Education next. Here's how*. Hacker Noon. https://hackernoon.com/blockchain-could-revolutionize-education-next-heres-how-b720bdf5945 [Accessed 12 Nov. 2022].

Sharples, M., & Domingue, J. (2016). The Blockchain and kudos: A distributed system for educational record, reputation and reward. *Adaptive and Adaptable Learning*, 490–496. https://doi.org/10.1007/978-3-319-45153-4_48.

Sheth, H. and Dattani, J., (2019). Overview of blockchain technology. *Asian Journal For Convergence In Technology (AJCT) ISSN-2350-1146*.

Shutikova, M. and Beshenkov, S., (2020). Modern digital educational environment and media education-platforms for transforming education system. *Медиаобразование*, *60*(4), pp.736-744.

Snyder, H., (2019). Literature review as a research methodology: An overview and guidelines. *Journal of business research*, *104*, pp.333-339.

Steiu, M.F., (2020). Blockchain in education: Opportunities, applications, and challenges. *First Monday*.

Stovall, D., (2005). A challenge to traditional theory: Critical race theory, African-American community organizers, and education. *Discourse: studies in the cultural politics of education*, *26*(1), pp.95-108.

Wallace, S. (2019). *The impact of blockchain technology on education*. [online] Available at: https://devm.io/blockchain/blockchain-education-161738 [Accessed 8 Nov. 2022].

Wegrzyn, K.E. and Wang, E. (2021). *Types of Blockchain: Public, Private, or Something in Between | Foley & Lardner LLP*. [online] www.foley.com. Available at: https://www.foley.com/en/insights/publications/2021/08/types-of-blockchain-public-private-between [Accessed 16 Nov. 2022].

World Bank. (2021). *Blockchain for Education Community of Practice*. [online] Available at: https://www.worldbank.org/en/topic/edutech/brief/blockchain-for-education-community [Accessed 13 Nov. 2022].

Yaga, D., Mell, P., Roby, N. and Scarfone, K., (2019). Blockchain technology overview. *arXiv preprint arXiv:1906.11078*.

Yli-Huumo, J., Ko, D., Choi, S., Park, S. and Smolander, K., (2016). Where is current research on blockchain technology?—a systematic review. *PloS one*, *11*(10), p.e0163477.